\documentstyle[12pt]{article}

\batchmode
  \newfont{\footscrfont}{rsfs10}
  \newfont{\footbbbfont}{msbm10}
\errorstopmode

\newif\ifscrf\scrftrue
\ifx\footscrfont\nullfont
  \scrffalse
\fi

\newif\ifamsf\amsftrue
\ifx\footbbbfont\nullfont
  \amsffalse
\fi

\def\ppnumber{\vbox{\baselineskip14pt\hbox{CLNS-95/1348}
\hbox{hep-th/9507012}}}
\def\ppdate{June 1995}
\def\pplogo{\vbox{\kern-\headheight\kern -15pt
\halign{##&##\hfil\cr&{
\ppnumber}\cr\rule{0pt}{2.5ex}&\ppdate\cr}
}}

\makeatletter
\date{}
\def\dedicatory#1{\def\@date{\normalsize\it#1}}
\def\subjclass#1{\def\@thefnmark{}\@footnotetext{1991
    {\it Mathematics Subject Classification.} #1}}
\def\keywords#1{\def\@thefnmark{}\@footnotetext{
    {\it Key words and phrases.} #1}}

\def\ps@firstpage{\ps@empty \def\@oddhead{\hss\pplogo}%
  \let\@evenhead\@oddhead 
}
\def\maketitle{\par
 \begingroup
 \def\thefootnote{\fnsymbol{footnote}}
 \def\@makefnmark{\hbox
 to 0pt{$^{\@thefnmark}$\hss}}
 \if@twocolumn
 \twocolumn[\@maketitle]
 \else \newpage
 \global\@topnum\z@ \@maketitle \fi\thispagestyle{firstpage}\@thanks
 \endgroup
 \setcounter{footnote}{0}
 \let\maketitle\relax
 \let\@maketitle\relax
 \gdef\@thanks{}\gdef\@author{}\gdef\@title{}\let\thanks\relax}

\def\abstract{\if@twocolumn
\section*{Abstract}
\else \small
\begin{center}
{\bf ABSTRACT}
\end{center}
\quotation
\fi}

\newif\iffn\fnfalse

\@ifundefined{reset@font}{\let\reset@font\empty}{} 
\long\def\@footnotetext#1{\insert\footins{\reset@font\footnotesize
    \interlinepenalty\interfootnotelinepenalty
    \splittopskip\footnotesep
    \splitmaxdepth \dp\strutbox \floatingpenalty \@MM
    \hsize\columnwidth \@parboxrestore
   \edef\@currentlabel{\csname p@footnote\endcsname\@thefnmark}\@makefntext
    {\rule{\z@}{\footnotesep}\ignorespaces
      \fntrue#1\fnfalse\strut}}}
\makeatother

\ifamsf
  \newfont{\bigbbbfont}{msbm10 scaled\magstep2}
  \newfont{\bbbfont}{msbm10 scaled\magstep1}  
  \newfont{\smallbbbfont}{msbm8}
  \newfont{\tinybbbfont}{msbm6}
  \newfont{\smallfootbbbfont}{msbm7}
  \newfont{\tinyfootbbbfont}{msbm5}
\fi

\ifscrf
  \newfont{\scrfont}{rsfs10 scaled\magstep1}  
  \newfont{\smallscrfont}{rsfs7}
  \newfont{\tinyscrfont}{rsfs7}
  \newfont{\smallfootscrfont}{rsfs7}
  \newfont{\tinyfootscrfont}{rsfs7}
\fi

\ifamsf
  \newcommand{\Bbb}[1]{\iffn
      \mathchoice{\mbox{\footbbbfont #1}}{\mbox{\footbbbfont #1}}
      {\mbox{\smallfootbbbfont #1}}{\mbox{\tinyfootbbbfont #1}}\else
      \mathchoice{\mbox{\bbbfont #1}}{\mbox{\bbbfont #1}}
      {\mbox{\smallbbbfont #1}}{\mbox{\tinybbbfont #1}}\fi}
\else
  \def\bigbbbfont{\bf}
  \def\Bbb{\bf}
\fi

\ifscrf
  \newcommand{\Scr}[1]{\iffn
    \mathchoice{\mbox{\footscrfont #1}}{\mbox{\footscrfont #1}}
    {\mbox{\smallfootscrfont #1}}{\mbox{\tinyfootscrfont #1}}\else
    \mathchoice{\mbox{\scrfont #1}}{\mbox{\scrfont #1}}
    {\mbox{\smallscrfont #1}}{\mbox{\tinyscrfont #1}}\fi}
\else
  \def\Scr{\cal}
\fi

\def\operatorname#1{\mathop{\rm #1}\nolimits}
\def\C{{\Bbb C}}

\def\P{{\Bbb P}}
\def\Q{{\Bbb Q}}
\def\R{{\Bbb R}}
\def\Z{{\Bbb Z}}

\def\Img{\operatorname{Im}}
\def\Rea{\operatorname{Re}}
\def\Gr{\operatorname{Gr}}
\def\rank{\operatorname{rank}}

\def\opeq#1{\advance\lineskip#1 \advance\baselineskip#1
	\advance\lineskiplimit#1}

\def\sm{$\sigma$-model}

\def\CY{Calabi-Yau}

\def\cM{{\Scr M}}

\def\cD{{\Scr D}}

\def\cMc{{\hfuzz=100cm\hbox to 0pt{$\;\overline{\phantom{X}}$}\cM}}
\def\barcD{{\hfuzz=100cm\hbox to 0pt{$\;\overline{\phantom{X}}$}\cD}}

\def\ff#1#2{{\textstyle\frac{#1}{#2}}}

\def\cMs#1{\cM_{\hbox{\scriptsize #1}}}

\ifamsf
  
  \def\rtimes{\mathbin{\mbox{\bbbfont\char"6F}}}
\else
  
  \def\rtimes{.}
\fi

\begin{document}
\setcounter{page}0
\title{\LARGE Enhanced Gauge Symmetries and K3 Surfaces\\[10mm]}
\author{
Paul S. Aspinwall\\[0.7cm]
\normalsize F.R.~Newman Lab.~of Nuclear Studies,\\
\normalsize Cornell University,\\
\normalsize Ithaca, NY 14853\\[10mm]
}

{\hfuzz=10cm\maketitle}

\def\Large{\large}
\def\LARGE{\large\bf}

\vskip 1.5cm
\vskip 1cm

\begin{abstract}

String-string duality dictates that type IIA strings compactified on a
K3 surface acquire non-abelian gauge groups for certain values of the
K3 moduli.  We argue that, contrary to expectation, the theories for
which such enhanced gauge symmetries appear are not orbifolds in the
string sense. For a specific example we show that a theory with
enhanced gauge symmetry and an orbifold theory have the same classical
K3 surface as a target space but the value of the ``B-field'' differs.
This raises the possibility that the conformal field theory associated
to a string theory with an enhanced gauge group is badly behaved in some
way.

\end{abstract}

\vfil\break

\section{Introduction}		\label{s:intro}

There is now strong evidence that the type IIA superstring
compactified on a K3 surface is equivalent to the heterotic string
compactified on a 4-torus
\cite{Sei:K3,AM:K3p,HT:unity,W:dyn,HS:sol,Sen:sol}. Both theories give
a theory of non-chiral $N=2$ supergravity in 6 dimensions and the
moduli spaces are identical. At a generic point in this moduli space
the gauge group in both theories is ${\cal G}\cong U(1)^{24}$. The
manifest difference in the world-sheet point of view of these theories
is explained by saying that the strong string-coupling limit of one is the
weak string-coupling limit of the other.

In the case of the heterotic string compactified on the 4-torus, it is
known that the gauge group in general can be ${\cal G}\cong {\cal G}_0
\times U(1)^4$, where ${\cal G}_0$ contains a semi-simple simply-laced
Lie group. This occurs for special points in the moduli space where
some vector states become massless. An interesting question is what
the corresponding statement for the type IIA string is. It is known that
from a world-sheet point of view, the gauge group must always be
abelian. Thus if this string-string duality is to work, the extra
massless vector particles in the type II theory must come from
non-perturbative effects, i.e., they are solitons.

Classically, as we will review below, one can show that the K3
surfaces giving such enhanced gauge symmetries for type II strings
have quotient singularities and are thus orbifolds as far as target
space geometry is concerned \cite{W:dyn}.  It will be important to
understand what is meant by the term ``orbifold'' and we will state it
here. Geometrically, an orbifold is a space whose only singularities
are locally in the form of some smooth space (usually $\C^n$) modded
out by some discrete group fixing the origin. A special case of such
an orbifold is the case of a global orbifold which may be written in
the form $Y/K$ for some smooth manifold, $Y$, admitting some group of
automorphisms, $K$.  In this global case, one can also define an
orbifold from a string theory, or conformal field theory, point of
view \cite{DHVW:} by directly taking the quotient of the field theory
describing $Y$ by the group $K$. It is well-known that such orbifold
theories are well-behaved despite the apparent target space being
singular.  Thanks to quantum geometry there is a subtle difference
between geometrical and string orbifolds which we will explain later.
It is probable that general features of string orbifolds can also be
extended to the case where the target space is only locally of the
form of a quotient \cite{me:orb2} so that it should be of little
importance whether an orbifold is global or not. In this letter, any
specific examples of orbifolds explicitly described will be of the
global type.

It is natural to speculate that the theories with enhanced gauge
symmetry are orbifolds from the conformal field theory point of view
as well as being orbifolds from the geometrical point of view. The
conformal field theory view of an orbifold is a perfectly well-behaved
theory and thus this enhancement of the gauge group would correspond
to solitons becoming massless when the underlying two-dimensional
point of view is smooth. This leads to surprising conclusions as
discussed in \cite{W:dyn}. This scenario would tell us, for example,
that we cannot trust the results of conformal field theory to give us
the correct massless spectrum even when the conformal field theory is
well-behaved. This should also be contrasted with the recent work of
\cite{Str:con} where solitons become massless at points in the moduli
space where the conformal field theory is singular.

We will show in this letter that, in a least one case, this classical
reasoning does not work and the point in moduli space representing the
conformal field theory orbifold and the point representing the theory
with enhanced gauge symmetry do not coincide --- they have a different
value of the ``$B$-field'' which cannot be observed using classical
geometry since it is the component of the $B$-field which is hidden
away in the quotient singularity. We believe this will be a general
phenomenon. The fact that the point of enhanced gauge symmetry is not
a string orbifold leaves open the possibility that the conformal field
theory description of this point is badly-behaved in some way.

In section \ref{s:het} we will analyze the enhanced gauge groups from
the heterotic string point of view. This allows us in section
\ref{s:K3m} to describe these points in terms of type II strings in K3
surfaces. We then use discrete symmetries in section \ref{s:sym} to
give an example of an orbifold which cannot have any enhanced gauge group.


\section{The Heterotic String} \label{s:het}

Let us first pin-point in the moduli space of heterotic string
theories where the enhanced gauge groups arise. The moduli space is of
the form \cite{N:torus}
\begin{equation}
  \cMs{H}\cong \Gr(\Pi,\R^{4,20})/O(\Lambda^{4,20}),
\end{equation}
where $\Pi$ is a space-like 4-plane and $\Gr(\Pi,\R^{4,20})$ denotes the
Grassmanian of such planes passing through the origin of $\R^{4,20}$
(often written $O(4,20)/(O(4)\times O(20))$). The lattice
$\Lambda^{4,20}$ is an even self-dual lattice of rank 24 embedded in
$\R^{4,20}$ and thus of the form $(E_8)^{\oplus2}\oplus H^{\oplus4}$,
where $E_8$ is {\em minus\/} the Cartan matrix of the Lie algebra $E_8$
and
\begin{equation}
  H = \left(\begin{array}{cc}0&1\\1&0\end{array}\right). \label{eq:H}
\end{equation}
The group $O(\Lambda^{4,20})$ is the automorphism group of this
lattice (often written $O(4,20;\Z)$).

The lattice $\Pi\cap\Lambda^{4,20}$ represents the set of right-moving
winding/momenta modes of the compactified heterotic string (before the
GSO projection) and the
lattice $\Pi^\perp\cap\Lambda^{4,20}$ represents the corresponding set of
left-moving states. Consider an element
$\alpha\in\Pi^\perp\cap\Lambda^{4,20}$ such that
\begin{equation}
  \alpha^2 = -2.   \label{eq:l=2}
\end{equation}
We will say that such an element has ``length'' equal to $-2$.
Such a state will be massless in the six-dimensional theory. In fact,
it will be a charged vector and it is not difficult to show that the
set of elements of $\alpha\in\Pi^\perp\cap\Lambda^{4,20}$ satisfying
(\ref{eq:l=2}) will form the set of roots for the semi-simple part of
the gauge group ${\cal G}_0$. Note that supersymmetry kills the
corresponding statement in the right-moving sector.

Thus we see that any simply-laced algebra that can be embedded in this
lattice can be realized by tilting $\Pi$ to be orthogonal to this
sub-lattice generated by the simple roots.


\section{Strings on K3 Surfaces}   \label{s:K3m}

Now let us review the form of the moduli space of type IIA strings
compactified on a K3 surface. This is equivalent to the moduli space of
$N$=4 superconformal field theories with K3 target space and is thus
of the form described in \cite{AM:K3p}. That is,
\begin{equation}
  \cMs{II}\cong \Gr(\Pi,\R^{4,20})/O(\Lambda^{4,20}),
\end{equation}
as required by string-string duality. To interpret this in terms of classical
geometry one decomposes $\Gr(\Pi,\R^{4,20})$ by the following
process.\footnote{This construction was made in collaboration with
D.~Morrison and was used in \cite{AM:K3p} although it was not
explicitly described there.} First choose a
primitive null-vector
$w\in\Lambda^{4,20}$. Now define $\Sigma^\prime=\Pi\cap w^\perp$. Then
define $B^\prime$ as a vector orthogonal to $\Sigma^\prime$ such that
$B^\prime.w=1$ and $\Pi$ is spanned by $\Sigma^\prime$ and $B^\prime$.

Consider now the subspace of $\R^{4,20}$ given by $w^\perp/w$. Such a
space is isomorphic to $\R^{3,19}$ and $\Lambda^{4,20}\cap(w^\perp/w)$
is an even self-dual lattice $\Lambda^{3,19}$. This space $w^\perp/w$
is taken to represent $H^2({\rm K3},\R)$ and the sublattice
$\Lambda^{3,19}$ gives the lattice of integral cohomology $H^2({\rm
K3},\Z)$. Projecting $\Sigma^\prime$ and $B^\prime$ into $w^\perp/w$
we define $\Sigma$ and $B$ respectively.

The space-like 3-plane $\Sigma\subset H^2({\rm K3},\R)$ describes a K3
surface, $X$, classically as follows \cite{Besse:E}. Let
$\Omega$ represent a
holomorphic 2-form on $X$ which is uniquely defined up to scale. It is
easy to then show that $\Rea(\Omega)$ and $\Img(\Omega)$ span a
space-like 2 plane in $H^2(X,\R)$. If $X$ is taken to have volume one
then the K\"ahler form, $J$, is an element of $H^2(X,\R)$ satisfying
$J^2=1$. $J$ is also a (1,1)-form and is thus orthogonal to
$\Rea(\Omega)$ and $\Img(\Omega)$. Thus $\Rea(\Omega)$, $\Img(\Omega)$
and $J$ span a space-like 3-plane. This plane is given by $\Sigma$ and the
Grassmanian of such planes in $\R^{3,19}$ gives precisely the
Teichm\"uller space for Ricci-flat metrics on a K3 surface of volume 1.

The vector $B$ is simple to interpret. It is the ``$B$-field'', $B\in
H^2(X,\R)$,  for the \sm. We have effectively used $w$ to give
a decomposition
\begin{equation}
  \Gr(\Pi,\R^{4,20}) \cong \Gr(\Pi,\R^{3,19})\times
\R^+\times\R^{3,19}.	\label{eq:decomp}
\end{equation}
This reads as ``factorizing'' the moduli space of type IIA strings on
a K3 surface into moduli of volume one K3 surfaces, the volume of the K3
surface and the value of the $B$-field respectively. The subgroup of
$O(\Lambda^{4,20})$ which preserves $w$ may act on the right of
equation (\ref{eq:decomp}) preserving the decomposition. The subgroup
$O(\Lambda^{3,19})$ acts on the first factor, as required classically,
and the subgroup of translations by $H^2(X,\Z)$ acts on the last factor as
expected from the nonlinear \sm.

Thus, given a point in the moduli space $\cMs{II}$ and a null vector
$w$, we may determine precisely which K3 surface with which $B$-field
our string theory is describing. Note that different choices of $w$
can lead to different geometrical interpretations allowing for stringy
equivalences between different K3 surfaces.

Now, given the results of section \ref{s:het}, let us determine which
K3 surfaces should have enhanced gauge symmetries. Let $L_{\cal G}$ be
the lattice generated by the simple roots of the gauge group. Then we
have
\begin{equation}
  L_{\cal G}\cong \Pi^\perp\cap \Lambda^{4,20}.
\end{equation}
Let us assume that we may choose $w$ such that $L_{\cal G}$ lies
within $w^\perp/w$ (which implies that $\rank L_{\cal G}\leq19)$. Since
the elements of $L_{\cal G}$ are orthogonal to $\Pi$, they are
orthogonal to $\Sigma^\prime$. There is a linear function $\phi:\Sigma
\to\R$ such that $x+\phi(x)w\in\Sigma^\prime$ for every
$x\in\Sigma$. Therefore, the elements of $L_{\cal G}$ are orthogonal
to $\Sigma$.

The elements of $L_{\cal G}$ are elements of $H^2(X,\Z)$ are are thus
dual to integral 2-cycles. Since they are orthogonal to $\Sigma$ they
are orthogonal to $\Rea(\Omega)$ and $\Img(\Omega)$ and are therefore
(1,1)-forms. Thus means that they are dual to algebraic curves in
$X$. For the elements of $L_{\cal G}$ with length $-2$, these
must be rational curves.
The area of these curves is given by the dot product with $J$
which is zero since $J$ lies in $\Sigma$.
Thus the simple roots of ${\cal G}_0$ give rational curves of zero
size in $X$. That is, $X$ is a blow-down of some smooth K3 surface
$\overline{X}$. The singularities of $X$ are thus locally of the form of a
quotient, i.e., {\em $X$ is an orbifold as far as geometry is
concerned.} This agrees with the expectation that the non-linear \sm\
must break down (due to singularities in the target space) in order
for non-perturbative effects to produce an enhanced gauge group. It
also nicely ties together the A-D-E classification
of surface quotient singularities with the possible simply-laced Lie groups
that can appear as gauge groups \cite{W:dyn}.

The analysis of \cite{W:dyn} is not quite complete for our purposes
however. It is important to note that the condition for an enhanced
gauge symmetry also puts constraints on the $B$-field. For $\Pi$ to be
orthogonal to any element $\alpha\in L_{\cal G}$, it is easy to show
from the above that $B.\alpha=0$. That is, the $B$-field must be zero
along certain directions. It follows that we may take $X$ to be an
orbifold and yet choose $B$ such that there is no enhanced gauge
symmetry by taking the values of $B$ to be nonzero in certain
directions.

One might expect that
the theories for which there is an enhanced gauge symmetry are
orbifolds from an conformal field theory point of view as well as
being orbifolds from the geometrical point of view. This makes a
statement about the value of $B$, as we now explain, that we will show
is not correct in at least one example.

Suppose we consider the familiar orbifold $T^4/\Z_2$. The
$\Z_2$-action has 16 fixed points resulting in 16 quotient
singularities in the orbifold. The orbifold may be blown-up to give a
K3 surface. Consider this model from the conformal field theory point
of view. We may divide the conformal field theory corresponding to
$T^4$ by the $\Z_2$ symmetry to obtain another conformal field theory
which is an orbifold. It is well-known that this orbifold theory is
well-behaved even though its apparent target space $T^4/\Z_2$ is
singular. Now consider the $B$-field degree of freedom. A K3 surface
has $b_2=22$ and so the $B$-field lives in a 22-dimensional space. The
4-torus has only 6 independent two-forms and thus the $B$-field in
this case lives in only a 6-dimensional space. Thus, when we make the
orbifold in the conformal field theory language we must be implicitly
assigning a value to the 16 components of the $B$-field coming from
the twisted sector. Only if this value is zero can we really assert
that our orbifold has an enhanced gauge group.

The value of the $B$-field for an orbifold was studied in
\cite{AGM:sd,me:orb2} for the case of nonlinear \sm s on complex
threefolds. There it was found that $B_i=\ff12$ for the twisted
component of the $B$-field. If we can make some kind of similar
statement for the case of complex surfaces we will have shown that
conformal field theory orbifolds never have enhanced gauge
groups. Actually, the case of complex surfaces is rather harder than
the threefold case. In the latter, it was the way that the $B$-field
appeared in the instanton expansion of the correlation functions of
the chiral primary fields that allowed it to be studied.  In the case
when the target space has complex dimension two it is simple to show
that such instanton effects are trivial. Such effects can be expressed
in terms of a topological field theory where the instantons are
rational curves \cite{W:AB}. In the case of complex dimension two the
target space has a quaternionic structure allowing any of an $S^2$ of
complex structures to be chosen for a given Ricci-flat metric. Whether
a rational curve exists or not depends on this choice of complex
structure. Since the field theory cares only about the Ricci-flat
metric and not the complex structure it would lead to a contradiction
if there were any nonzero instanton effects.  We therefore need some
other way to determine $B$. This will be the use of discrete
symmetries as we now discuss in the following section.


\section{K3 Surfaces with Symmetries}  \label{s:sym}

Consider a K3 surface $X$, with a finite group, $G$, of
automorphisms which preserve the holomorphic 2-form, $\Omega$. Such
symmetry groups, $G$, are well-understood thanks to the beautiful work
of Nikulin \cite{Nik:K3aut} for the abelian case and Mukai
\cite{Muk:K3aut} in the general case.

The general question we wish to address is whether we can find an
example of a K3 surface with such a large symmetry group that its
position in the moduli space is uniquely determined. Let us first
analyze the classical case. To do this we introduce 4 sublattices of
$H^2(X,\Z)\cong\Lambda^{3,19}$:
\begin{enumerate}
\item $S_X$ is the lattice of {\em algebraic cycles}, i.e., elements of
$H^2(X,\Z)$ orthogonal to $\Rea(\Omega)$ and $\Img(\Omega)$.
\item $T_X$ is the lattice of {\em transcendental cycles}, i.e.,
$T_X=S_X^\perp\subset H^2(X,\Z)$.
\item $T_{X,G}$ is the lattice of $G$-invariant elements of
$H^2(X,\Z)$.
\item $S_{X,G}=T_{X,G}^\perp\subset H^2(X,\Z)$.
\end{enumerate}
We then have the following Lemma (4.2 of \cite{Nik:K3aut}):
\begin{enumerate}
\item[a)] $S_{X,G}$ is nondegenerate and negative definite.
\item[b)] $S_{X,G}$ does not contain any elements, $\alpha$, such that
$\alpha^2=-2$.
\item[c)] $T_X\subset T_{X,G}$ and $S_{X,G}\subset S_X$.
\end{enumerate}

If we want the automorphism to preserve the K\"ahler form, $J$, as
well as the holomorphic 2-form we require that the 3-plane $\Sigma$
spanned by $J$ and $\Omega$ lies in $T_{X,G}\otimes\R$. Thus the rank of
$S_{X,G}$ can be no larger than 19. If it is 19 then $\Sigma$ is fixed
uniquely by $G$.

In \cite{Muk:K3aut} the space of total rational cohomology was
introduced, $V=H^*(X,\Q)$. The quantity $\mu(G)$ was introduced, where
\begin{equation}
\mu(g) = \frac{24}{n\displaystyle\prod_{p|n}\left(1+\frac1p\right)},
\end{equation}
where $g\in G$, $n$ is the order of $g$, the product runs over the
prime factors of $n$ and
\begin{equation}
\mu(G) = \frac1{|G|}\sum_{g\in G}\mu(g).
\end{equation}
One may then show that $\mu(G)$ is equal to the dimension of the
$G$-invariant part of $V$. Since $H^0$ and $H^4$ are clearly preserved
by $G$ we have that $\mu(G)\geq5$ with equality fixing $\Sigma$
uniquely.

Consider the case where $X$ is the ``Fermat'' quartic K3 surface
defined as the hypersurface
\begin{equation}
  x_0^4+x_1^4+x_2^4+x_3^4=0\quad\subset\P^3.
\end{equation}
This has symmetries generated by multiplying the homogeneous
coordinates, $x_j$, by $i$ and by permuting the homogeneous
coordinates. All told, we have $G\cong((\Z_4)^3\rtimes{\cal
S}_4)/\Z_4$, where ${\cal S}_4$ is the symmetric group on 4 elements.
An elementary calculation shows that $\mu(G)=5$ and so we have an
example where the symmetry is maximal in that $S_{X,G}$ is rank 19 and
$\Sigma$ is uniquely fixed.

Let us note at this point the significance of the fact that $S_{X,G}$
contains no elements of length $-2$. From our discussion in section
\ref{s:K3m} we see that, since $\Sigma$ is orthogonal to elements in
$S_{X,G}$, an element of length $-2$ would give a quotient singularity
in $X$. Thus $X$ cannot be an orbifold. In this case it is obvious ---
the Fermat quartic surface is indeed smooth --- but we see that having
a large symmetry group itself prevents $X$ acquiring singularities. It is
this statement that we now wish to extend to the quantum case.

\def\wh{\widehat}

Let us extend the above to the case of stringy geometry along the lines
suggested in \cite{AM:K3p}. Now we take the lattice
$H^*(X,\Z)\cong\Lambda^{4,20}$ and allow $G$ to act on that. We define
lattices in analogy with above, namely let $\wh T_{X,G}$ be the
sublattice of $H^*(X,\Z)$ preserved by $G$, and $\wh S_{X,G}$ be the
orthogonal complement in $H^*(X,\Z)$. The analogy of the statement
that $G$ preserves $\Omega$ in the quantum case is that $G$ preserves
the chiral primary field of charge $(2,0)$ of the conformal field
theory. In this case $\Pi\subset\wh T_{X,G}\otimes\R$. Thus $\wh S_{X,G}$ has
rank $\leq 20$.

This may be related to the classical case as follows. Choose a
primitive null vector $w$ as in section \ref{s:K3m} to fix the
geometrical interpretation of the theory. We then claim that $G_c$,
the group of classical symmetries, is the subgroup of $G$ that fixes
$w$ and that
\begin{equation}
  S_{X,G_c} = \wh S_{X,G}\cap w^\perp/w.
\end{equation}
This means that by choosing various $w$'s we may extend Nikulin's
lemma to the quantum case. In particular $\wh S_{X,G}$ is negative
definite and has no elements of length $-2$.

If we can find an example where $\wh S_{X,G}$ is of rank 20 then this
will fix $\Pi$ uniquely. We can do this by taking a case where the
classical K3 surface has $\rank S_{X,G_c}=19$ and there is a ``quantum
symmetry'' present in addition. A quantum symmetry is a symmetry of
the string theory but not of the classical geometry and so must be an
element of $G$ that does not preserve $w$.

An example of this is found by taking the Gepner model \cite{Gep:} of the
Fermat quartic K3 surface. This model amounts to taking a particular
value for the K\"ahler form and $B$-field and it has a non-trivial quantum
symmetry as described in \cite{Vafa:qu}. (The extra $\Z_4$ which
appears at this point does not itself preserve the (2,0)-field but one
can mix this with other classical symmetries present to
produce one that does.) It follows that {\em the quartic K3
surface at the Gepner point is a model with $\rank \wh S_{X,G}=20$.}
In particular, since $\Pi^\perp\cap\Lambda^{4,20}$ is $\wh S_{X,G}$
and so has no elements of length $-2$, by string-string duality we
infer that {\em this model has no enhanced gauge symmetry}.

It now remains to show that this model has an orbifold
interpretation. This is actually quite easy given the Greene-Plesser
mirror construction \cite{GP:orb}. The mirror of this quartic K3
surface, $X$ by this construction is the orbifold
$X/(\Z_4\times\Z_4)$. Type IIA strings compactified on a K3 surface
are self-mirror \cite{AM:Ud} and so this orbifold theory is just a
different geometrical interpretation of the same string theory --- we
have just chosen another $w$. Thus we have given an example of an
orbifold conformal field theory with no enhanced gauge symmetry. This
means that the effective value of the $B$-field is nonzero along
certain twisted directions. Analyzing the fixed-point set of this
orbifold one sees that the gauge group will be ${\cal G}\cong SU(4)^6
\times U(1)^6$ if we move a finite distance in the moduli space by
integrating along the twisted
marginal operators that will take the value of the $B$-field to zero.


\section{Conclusions}

Classically, for the case of type IIA strings on a K3 surface, the
enhanced gauge groups appear when the underlying space is an orbifold
but we have shown here that quantum geometry can move this point away
from the conformal field theory orbifold point in the moduli space of
string theories.  That is, we have given an example of an orbifold in
the string (or conformal field theory) sense which has no enhanced
gauge symmetry.  Suppose, for a moment, that we could find a string
orbifold which did have an enhanced gauge group when used to
compactify the type IIA string.  This would mean that the conformal
field theory approach was unable to compute the massless spectrum of
states despite being apparently well-behaved.  This is to be
contrasted with the situation in \cite{Str:con} where massless
solitons coincided with badly behaved conformal field theories.  It
would be interesting to see if the conformal field theory associated
with the enhanced gauge symmetry point is bad in some way.  Indeed, in
section 5.5 of \cite{AGM:sd} an example was given in the threefold
case of an orbifold theory which becomes singular when the $B$-field
is taken to zero.  If one can understand such effects in the case of
two complex dimensions, this may provide a method of generalizing the
results of this letter.


\section*{Acknowledgements}

It is a pleasure to thank M. Gross, D. Morrison, S. Shenker, H. Tye
and E. Witten
for useful conversations.
The work of the author is supported by a grant from the National
Science Foundation.


\end{document}